\documentclass[aps,prb,twocolumn,amsmath,amssymb,nofootinbib,superscriptaddress,floatfix,eqsecnum]{revtex4-1}

\usepackage{amsmath}
\usepackage{amssymb}
\usepackage{amsthm}
\usepackage[dvipdf]{color}
\usepackage{graphicx}
\usepackage{dcolumn} 
\usepackage{bm} 
\usepackage[hypertex]{hyperref}
\usepackage{srcltx}

\begin{document}

\title{
Interacting topological phases and modular invariance
}

\author{Shinsei Ryu}
\affiliation{
Department of Physics,
University of Illinois at Urbana-Champaign,
1110 West Green St, Urbana,
Illinois 61801, USA
            }

\author{Shou-Cheng Zhang}
\affiliation{
Department of Physics, Stanford University,
Stanford, California 94305, USA
            }

\date{\today}

\begin{abstract}
We discuss 
(2+1)-dimensional topological superconductors
with $N_f$ left- and right-moving Majorana edge
modes and a $\mathbb{Z}_2\times \mathbb{Z}_2$ symmetry.
In the absence of interactions, these phases are distinguished 
by an integral topological invariant $N_f$. With interactions,
the edge state in the case of $N_f=8$ is unstable against 
interactions, and a $\mathbb{Z}_2\times \mathbb{Z}_2$ invariant
mass gap can be generated dynamically. 
We show that this phenomenon
is closely related to the modular invariance of 
type II superstring theory. 
More generally, we show that the global gravitational 
anomaly of the nonchiral Majorana edge states is the physical 
manifestation of the bulk topological superconductors classified 
by $\mathbb{Z}_8$.
\end{abstract}

\maketitle

\section{Introduction}

Topological insulators and superconductors
are a gapped phase of matter
with a stable gapless mode at their boundary.
A classic example is the integer quantum Hall effect (IQHE),
which exists for two spatial dimensions
in the presence of
a strong time-reversal symmetry breaking magnetic field.
\cite{review_QHE}
A flurry of recent excitement
came with the discovery
of topological insulators
in two and three dimensions
in systems
in the presence of strong spin orbit coupling. 
\cite{review_TIa,review_TIb,KaneMele,Bernevig05,Bernevig06,Moore06,Roy3d,Fu06_3Da,Fu06_3Db,qilong}
Unlike the IQHE,
the topological character of these topological insulators
({\it i.e.}, the stable gapless edge or surface modes)
is protected by time-reversal symmetry (TRS).
With a wider set of discrete symmetries
in addition to TRS,
such as particle-hole symmetries of various kinds
realized in insulating and superconducting systems,
one can
ask if there is a topological distinction
among gapped phases in the presence of such symmetries.
The answer to this question is summarized in
the systematic classification of topological insulators and
superconductors.
\cite{qilong,Schnyder2008,SRFLnewJphys,Kitaev2009}

While these non-interacting topological phases
are stable against arbitrary deformation of the Hamiltonian
at the quadratic level, they could be fragile against fermion 
interactions. In the case of 
three-dimensional topological insulators, the
topological invariant can be physically defined in terms of
the topological magneto-electric effect with a quantized
coefficient,
\cite{qilong} 
which can be evaluated for a
generally interacting system in terms of the many-body
Green's function.
\cite{Wang2010} For this reason, we can
expect topological insulators to be stable against a general
class of interactions. However, Refs.\
\onlinecite{FidkowskiKitaev2009,FidkowskiKitaev2010,TurnerPollmannBerg2010}
also provided counter-examples in the case of topological superconductors. 
It was demonstrated that in (1+1) dimensional lattice Majorana fermion models,
with a suitable choice of interactions,
one can find an adiabatic path that connects
what appears to be a topological phase at the quadratic level
and a topologically trivial phase.

In this paper, we discuss a (2+1)-dimensional topological superconductor
with $N_f$ left- and right-moving Majorana edge
modes, and a $\mathbb{Z}_2\times \mathbb{Z}_2$ symmetry between them
(see Sec.\ \ref{Z2xZ2 symmetric topological phases}).
The similar/same models were studied independently 
in Refs.\ \onlinecite{Gu12,Qi12,YaoRyu12,GuLevin}.
In the absence of interactions,
these phases are distinguished by an integral
topological invariant, since they support
an integral number of non-chiral edge modes ($=N_f$).
With interactions, the edge state of the phase with $N_f=8$ is 
unstable to interactions. Therefore, the interacting phases of 
this model are classified by the $\mathbb{Z}_8$ topological
class (Sec.\ \ref{Z2xZ2 symmetric topological phases}).
We argue that this phenomenon is closely
related to the modular invariance of 
type II superstring
(Sec.\ \ref{edge theory of Z2xZ2 symmetric topological phase}).
More generally, we show that the global gravitational 
anomaly or the modular non-invariance of the non-chiral Majorana
edge states is the physical manifestation of the (2+1) bulk topological
superconductor (Sec.\ \ref{global gravitational anomaly}).

\section{$\mathbb{Z}_2\times \mathbb{Z}_2$ symmetric
topological phases}
\label{Z2xZ2 symmetric topological phases}

\subsection{Description of the model}

The topological phases of our interest
are in (2+1) dimensions,
and have
$\mathbb{Z}_2 \times \mathbb{Z}_2$
symmetry
with two
conserved $\mathbb{Z}_2$ quantum numbers.
A convenient way to describe these
quantum numbers is to first consider
systems with two conserved U(1) charges,
and then later break
the $\mathrm{U}(1)\times \mathrm{U}(1)$
symmetry
down to
$\mathbb{Z}_2 \times \mathbb{Z}_2$.
The two charges can be thought of as
the total fermion number
and
the total $S_z$
(the $z$ component of spin-1/2 operator)
quantum number,
denoted by
$N_{\uparrow}+N_{\downarrow}$,
and
$N_{\uparrow}-N_{\downarrow}$,
respectively.
We break the particle number conservation
by introducing superconducting pair potential,
so the system belongs to the Bogoliubov-de Genne (BdG)
symmetry class (class D) of Altland-Zirnbauer.
Here, we deal with the pairing potential
at the mean field level, and regard it simply as a
background.
In effect, we are considering quadratic Hamiltonians of
real fermions (BdG quasiparticles)
instead of complex fermions.
The pair potential breaks the electromagnetic U(1) symmetry,
and the total fermion number $N_{\uparrow}+N_{\downarrow}$ 
is now conserved only modulo 2, {\it i.e.}, the total fermion 
number parity $(-1)^{N_{\uparrow}+N_{\downarrow}}$ is conserved.

When the total $S_z$ is conserved,
the BdG Hamiltonians
can be block-diagonalized in the basis where $S_z$ is diagonal
(each block in the BdG Hamiltonians is a member of symmetry class A).
We now relax the conservation of total $S_z$,
and demand only the parity
$(-1)^{N_{\uparrow}}$
[or $(-1)^{N_{\downarrow}}$]
to be conserved;
combined with the total fermion number parity
conservation, 
the systems of our interest conserve two
$\mathbb{Z}_2$ quantum numbers,
$(-1)^{N_{\uparrow}}$
and
$(-1)^{N_{\downarrow}}$.
Even without strict conservation of $S_z$,
at the quadratic level, the BdG Hamiltonians
still remain block-diagonal since
the $\mathbb{Z}_2 \times \mathbb{Z}_2$ 
symmetry 
does not allow any spin flip,
{\it i.e.}, any bilinear connecting spin up and spin down sectors.
(So far, relaxing the $S_z$ conservation down
to the conservation of the two $\mathbb{Z}_2$ quantum numbers
does not change the story much at the quadratic level,
but it will make a big difference when we talk about interactions.)

These sub blocks in the BdG Hamiltonians
belong to symmetry class A (the same symmetry class as IQHE)
and their topological character is specified by
the Chern number,
$\mathrm{Ch}_{\uparrow}$ and  $\mathrm{Ch}_{\downarrow}$,
respectively;
the topological classes of the system is characterized by
a $\mathbb{Z}\times \mathbb{Z}$ topological number.

When
$\mathrm{Ch}_{\uparrow} + \mathrm{Ch}_{\downarrow} \neq 0$,
time-reversal symmetry (TRS) is necessarily broken,
and
a time-reversal symmetry breaking topological superconductor
(in symmetry class D) is realized.
This phase has non-zero thermal Hall conductance
$\kappa_{xy}$,
and
when there is an edge, it supports
an integer number ($=\mathrm{Ch}_{\uparrow} + \mathrm{Ch}_{\downarrow}$)
of {\it chiral} Majorana fermions.
This phase is robust against interactions as well as disorder.

The phase of our interest in this paper corresponds to the case with
the vanishing total Chern number,
$\mathrm{Ch}_{\uparrow} + \mathrm{Ch}_{\downarrow} = 0$
(this is guaranteed when there is time-reversal symmetry),
but with the non-zero spin Chern number,
$\mathrm{Ch}_s :=
(\mathrm{Ch}_{\uparrow} - \mathrm{Ch}_{\downarrow})/2 \neq 0$.
A lattice model that realizes this situation
can easily be constructed, by combining two
copies of lattice chiral $p$-wave superconductors
with opposite chiralities.
(See, for example, Ref.\ \onlinecite{YaoRyu12}.)
Similarly to the case of
$\mathrm{Ch}_{\uparrow} + \mathrm{Ch}_{\downarrow} \neq 0$,
the phase with $\mathrm{Ch}_s \neq 0$
supports edge modes
but unlike the case of $\mathrm{Ch}_{\uparrow} + \mathrm{Ch}_{\downarrow} \neq 0$,
edge modes are {\it non-chiral}.
Below, we will have a closer look
at the edge modes.

Let us begin with the case of $\mathrm{Ch}_s =1$.
The edge of the system
consists of a single copy of Majorana fermion with both
left- and right-moving chiralities,
described by the following Euclidean Lagrangian:
\begin{align}
\mathcal{L}
&=
\frac{1}{4\pi}
\big[
\psi_L (\partial_{\tau} + {i}v \partial_x) \psi_L
+
\psi_R (\partial_{\tau} - {i}v \partial_x) \psi_R
\big],
\end{align}
where $\tau$ is the imaginary time
and $x$ is the spatial coordinate parameterizing the edge;
$\psi_L$ ($\psi_R$) is the left- (right-) moving
(1+1) Majorana fermion, and $v$ is the Fermi velocity.
Here, one could think of
the left-mover to carry ``spin up''
and the right-mover to carry ``spin down'' quantum numbers,
respectively
(or vice versa, depending on the sign of $\mathrm{Ch}_s$).
As emphasized before, however,
we do not require the $S_z$ quantum number
to be conserved
[$N_{\uparrow}$ (or $N_{\downarrow}$) is conserved only up to modulo 2].
This means, in particular,
we do not have well-defined spin Hall conductance
$\sigma^s_{xy}$.
More generically, when $\mathrm{Ch}_s =N_f$,
the edge is described by $N_f$-flavor of Majorana fermions
with both left- and right-moving chiralities:
\begin{align}
 \mathcal{L}
&=
\frac{1}{4\pi}
\sum^{N_f}_{a=1}
\big[
\psi^a_L (\partial_{\tau} + {i}v \partial_x) \psi^a_L
+
\psi^a_R (\partial_{\tau} - {i} v \partial_x) \psi^a_R
\big].
\label{edge theory N=8}
\end{align}

Since they are non-chiral,
the gapless nature of the edge modes are not
stable in the absence of any symmetry; one can find a suitable
mass term that opens a gap.
Since the bulk of the system respects
$\mathbb{Z}_2\times \mathbb{Z}_2$ symmetry,
this is inherited by the edge theory;
we define two fermion parities in the edge theory,
\begin{align}
G_L = (-)^{N_L}
\quad
\mbox{and}
\quad
G_R = (-)^{N_R},
\end{align}
where $N_L(=N_{\uparrow})$ [$N_R(=N_{\downarrow})$] is the
total left-moving (right-moving) fermion numbers
at the edge.
With the $\mathbb{Z}_2\times\mathbb{Z}_2$ symmetry,
at the quadratic level,
all mass terms
$\psi^a_{L} \psi^b_{R}$
are prohibited as they are odd under
the left- or right-$\mathbb{Z}_2$ parity
($G_L$ or $G_R$)
--
bulk topological phase is characterized by an integer,
which is simply the number of branches of the (non-chiral) modes,
$N_f$.

\subsection{Effects of interactions}

Beyond the quadratic level, we can write down interactions
$\psi^a_{L} \psi^b_{L}\psi^c_{R}\psi^d_{R}$
that preserve $\mathbb{Z}_2\times \mathbb{Z}_2$
symmetry.
The presence of such interactions can potentially destabilize
the edge.
\cite{Wu2006} 
However, one would expect that the resulting gapped phase
would spontaneously break
$\mathbb{Z}_2\times \mathbb{Z}_2$;
at the mean-field level, such interactions
generate the expectation value
$\langle \psi^a_{L}\psi^b_{R} \rangle \neq 0$
for some pair of flavor indices $(a,b)$,
and if so $\mathbb{Z}_2\times \mathbb{Z}_2$ conservation is violated.

When $N_f=8$ (more precisely, when $N_f\equiv 0$ mod 8),
there is another type of interaction channel available
that can potentially destabilize the edge --
interactions in terms of ``spin'' or ``disorder'' operators.
Let us first recall the case of $N_f=1$,
the Ising conformal field theory (CFT).
In the quantum Ising model, we have two relevant operators;
the transverse field, and the Zeeman field.
The former, in the language of the two-dimensional classical Ising model,
corresponds to
the deviation from the critical temperature $(T-T_c)$,
and is given by the fermion mass term $\psi_L\psi_R$.
The latter, the Zeeman field,
while it is a natural and local perturbation in terms of
the Ising spin variable,
is a non-local term when the model is viewed as a fermion model.
This is so because of the Jordan-Wigner string.
In fact,
the operator product expansion between the Majorana fermion
$\psi_{L,R}$
and
the spin operator $\sigma_{L,R}$
has a branch cut,
signaling they are not a local object in terms of fermions.
In fact, the spin operator is a twist operator for the fermion 
field $\psi_{L,R}$; when $\sigma$ is inserted, say, at the origin,
when $\psi$ makes a round trip around the origin, it picks
up a minus sign.

When $N_f=8$,
from spin and disorder operators,
we can form $2^8=256$ possible products of
$\sigma^a(z,\bar{z})$ and
$\mu^a(z,\bar{z})$
($a=1,\ldots,N_f$).
These have conformal weight
$(1/16,1/16)\times 8=(1/2,1/2)$,
which is the conformal weight of
free fermions.
These fermions, which are different from
the original fermions $\psi^a_{L,R}$,
can be used to form a perturbation to the
edge theory, which are local with respect to
$\psi^a_{L,R}$. This is rooted in the triality symmetry of
$\mathrm{SO}(8)$. 
\cite{Shankar}
Assuming, for simplicity, that all eight
Majorana fermions $\psi^a_{L,R}$
have the same
Fermi velocity, the kinetic term of the edge theory
enjoys $\mathrm{SO}(8)$ symmetry.
The Majorana fermions $\psi^a_{L/R}$ belong to the vector representation
of $\mathrm{SO}(8)$,
$\mathbf{8}_v$. 
For $\mathrm{SO}(8)$, by ``accident'',
spinor ($\xi$) and conjugate spinor ($\eta$) are also eight dimensional
(denoted by $\mathbf{8}_s$ and $\mathbf{8}_c$, respectively),
the triality symmetry permutes these three representations.
The $2^8=256$ possible products of
$\sigma^a(z,\bar{z})$ and
$\mu^a(z,\bar{z})$
are precisely the (linear combination of)
$64\times 4$ primary fields
$\xi^a_R\xi^b_L$,
$\xi^a_R\eta^b_L$,
$\eta^a_R\xi^b_L$,
$\eta^a_R\eta^b_L$.
\cite{footnoteSO8}
These $\mathrm{SO}(8)$ spinors
can be described in terms of
Abelian bosonization as follows:
we pair up the vector fermions,
and bosonize as
\begin{align}
\psi^{2j-1}_L \pm {i} \psi^{2j}_L
&\simeq
\exp( \pm {i} \varphi^j_{L}),
\nonumber \\
\psi^{2j-1}_R \pm {i} \psi^{2j}_R
&\simeq
\exp( \pm {i} \varphi^j_{R}),
\end{align}
($j=1,\ldots, 4$). 
The $16=8+8$ fields
\begin{align}
\exp
\frac{ {i} }{2}
\left(
\pm \varphi^1_{L}
\pm \varphi^2_{L}
\pm \varphi^3_{L}
\pm \varphi^4_{L}
\right)
\end{align}
are the spinor $\xi^a_L$ and $\eta^a_L$,
with $\mathbb{Z}_2$ parity determined by
the parity of the number of minus signs in the 
exponential. $\xi^a_L$ is even under 
$\mathbb{Z}_2$ parity, while $\eta^a_L$
and $\psi^a_L$ are odd under
$\mathbb{Z}_2$ parity.

Since $\xi^a_L$ and $\xi^a_R$ are even under the
$\mathbb{Z}_2\times \mathbb{Z}_2$ parity, it is now
possible to construct quadratic terms $\xi^a_L \xi^b_R$
that could gap the edge states without violating
the $\mathbb{Z}_2\times \mathbb{Z}_2$ symmetry.
We use the interaction term constructed
in Ref.\ \onlinecite{FidkowskiKitaev2009},
which is given by
the Euclidean Lagrangian
\begin{align}
\mathcal{L}_{\mathrm{int}}
& =
-A
\left(
\sum\nolimits^7_{a=1} 
\xi^a_L \xi^a_R
\right)^2
- B
\left(
\sum\nolimits^7_{a=1} \xi^a_L \xi^a_R
\right)
\xi^8_L \xi^8_R,
\label{FK interaction}
\end{align}
where $A$ and $B$ are some constant.
This interaction can, in fact, also be expressed in terms of
the vector fermions $\psi^a_{L/R}$ because of triality,
and hence be a local interaction.
The SO(8) symmetry is broken down to SO(7)
which leaves the spinor $\xi^8_{L/R}$ invariant.

This interaction,
when $B<0$ and $B<2A$,
gives rise to a unique ground state
as we can see as follows:
when $B\ll A$,
because of the dominant SO(7) Gross-Neveu interaction term
$-A \big(\sum^7_{a=1} \xi^a_L \xi^a_R\big)^2$,
the bilinear $\sum^7_{a=1} \xi^a_L \xi^a_R$
develops an expectation value
$\big\langle \sum^7_{a=1} \xi^a_L \xi^a_R
\big\rangle={i} M$.
The interaction can then behave effectively as a mass term for
$\xi^8_{L/R}$,
$
\mathcal{L}_{\mathrm{int}}
\simeq
-{i} B M
\xi^8_L \xi^8_R$.
Thus, when $B\ll A$,
the model behaves essentially as
a single copy of the Ising model.
Depending on the sign of the induced mass $-B M$,
it can be either in
the low-temperature (symmetry broken)
or
the higher-temperature (paramagnetic) phase.
To determine which phase is realized,
we first note that when $B=2A$,
the interaction term is the SO(8) Gross-Neveu interaction.
This then leads to
a gapped phase with two-fold degenerate ground states
because of chiral symmetry breaking.
We would then conclude that when $B\ll A$ and $B>0$
(and in fact, for the entire region of $B> 0$ and $B>-2A$),
the model is in the low-temperature phase of
the effective Ising model, with two-fold degenerate ground states.
Next, we note that the sign of $B$ can be flipped in
the interaction (\ref{FK interaction})
by $\xi^a_R \to -\xi^a_R$,
the Kramers-Wannier duality transformation.
Thus, we conclude,
when $B\ll A$ and $B<0$
(and in fact, for the entire region of $B<0$ and $B<2A$),
the effective Ising model is in
the high-temperature phase (paramagnetic phase)
with unique ground state.
It can be checked that
the ground state does not violate
the $\mathbb{Z}_2\times \mathbb{Z}_2$ symmetry.

The discussion above can be formulated in a language
more familiar in the context of correlated electron
systems. When $N_f=8$, the eight Majorana fermions
can be mapped onto four complex fermions of a two-leg
ladder
(see, for example, Refs.\ 
\onlinecite{Scalapino1998, Lin1998,Marston2002, Fjaerestad2002,
Tsuchiizu2002,Schollwock2002,WuLiuFradkin2003},
and references therein)
or the spin-$3/2$ Hubbard model 
\cite{Wu2003}, with a suitable choice of
basis states. Interactions of a
two-leg ladder can be described by the on-site Hubbard
interaction $U$, the rung interaction $V$ and the rung
exchange $J$. When $J=4(U+V)$, the model is $\mathrm{SO(5)}$
symmetric at half-filling. 
\cite{Scalapino1998}
Furthermore, when $V=0$ or $J=4U$, the model is also
$\mathrm{SO(7)}$ symmetric, which in a suitable basis can also
be expressed as Eq.\ (\ref{FK interaction}). This interaction
can either lead to a unique rung singlet ground state,
or a two-fold degenerate staggered flux ground states.
\cite{Tsuchiizu2002} The quantum phase transition
between these states can be described by the transverse
field Ising model, 
\cite{Tsuchiizu2002} 
or equivalently,
by a single Majorana spinor, which is nothing but our
spinor $\xi^8$. In this sense, the high temperature or
the paramagnetic phase of the $\xi^8$ spinor corresponds
to the rung singlet state of a two-leg ladder, with a
gap generated by interactions.
\cite{latticevscont}

Alternatively,
one can postulate an interaction
that is $\mathbb{Z}_2\times \mathbb{Z}_2$ symmetric,
and involves both spinors and conjugate spinors,
\begin{align}
\mathcal{L}'_{\mathrm{int}}
& =
-A
\left(
\sum\nolimits^7_{a=1} \xi^a_L \eta^a_R
\right)^2
-
B
\left(
\sum\nolimits^7_{a=1} \xi^a_L \eta^a_R
\right)
\xi^8_L \eta^8_R.
\label{FK interaction 2}
\end{align}
Following the same reasoning,
this interaction gives rise to,
when $B<0$ and $B<2A$, a unique ground state.

From these discussion,
we conclude that the
$\mathbb{Z}_2\times \mathbb{Z}_2$ symmetric topological phases,
while it can support an integer number of non-chiral edge modes
when non-interacting, interactions make them unstable when $N_f=8$.
Therefore, interacting models falls into $\mathbb{Z}_8$ topological
classes. 
In the following sections, we will look more into the
reasons behind this stability/instability.

\section{Global gravitational anomaly}
\label{global gravitational anomaly}

\subsection{Large gauge transformations in electromagnetism}
\label{large gauge transformations in electromagnetism}

Our analysis on the stability/instability of the topological phases
so far relies on an explicit construction of an interaction
term in terms of the twist (spin and disorder) operators.
For the QHE and for the quantum spin Hall effect (QSHE), however,
their stability (and also instability in the case of the QSHE)
against interactions
can be understood from a wider
(more ``topological'')
point of view;
\cite{review_QHE,qi2008,LevinStern2009}
it is the Laughlin's thought experiment (and its suitable
extension to the QSHE),
which we will review briefly below
for our later discussion.
For our situation, since
the particle number and $S_z$ quantum number are not conserved
(conserved only mod 2),
we cannot rely on the flux(es) of U(1) gauge field
of charge or spin origin.
We will, instead, try to make use of gravitational field.

Let us consider the QHE on a finite cylinder
(which is topologically equivalent to an annulus). 
There are two edges, which we call ``edge I'' and ``edge II''.
We thread a magnetic flux $\Phi$ into the ``hole'' of the cylinder.
Starting from zero flux, let us gradually increase the flux.
The Hamiltonian $H(\Phi)$ of the system,
when $\Phi$ is not an integer multiple of the flux quantum $\Phi_0$,
is not gauge equivalent to
the original Hamiltonian; the insertion of the flux
is a physically effect, and not a gauge transformation.
When flux is an integer multiple of flux quantum, however,
the Hamiltonian goes back to itself,
$H(\Phi)=H(\Phi+ n \Phi_0)$ ($n\in \mathbb{Z}$).
This is an example of {\it large} gauge transformations;
the Hamiltonian with $n$ extra flux quanta
$n \Phi_0$ cannot be generated from
the original flux $\Phi$ by a successive application of
infinitesimal gauge transformation.
Unlike an infinitesimal gauge transformation,
to achieve such gauge transformation
by an adiabatic process,
one needs to generate physical flux during the process.

The same is true for the total partition function
$Z$ of the system as a function of flux $\Phi$:
it is invariant
under a large gauge transformation $\Phi\to \Phi + n \Phi_0$,
\begin{align}
Z(\Phi)
=
Z(\Phi+n \Phi_0).
\end{align}
However, in the QHE,
a closer inspection tells us that
in the adiabatic process where we increase
the flux from
$\Phi$
to
$\Phi + \Phi_0$, say,
an integer multiple of charge
is pumped from edge I to edge II (or edge II to edge I).
This means, if we focus on a single edge (edge I or edge II),
instead of the combined system of the two edges,
it looks as if the charge is not conserved.

Since the bulk is fully gapped,
for adiabatic processes,
it is meaningful to focus on excitations at the edges,
neglecting gapped excitations in the bulk.
The total partition function can then be written as
\begin{align}
Z (\Phi)
=
\sum_{a, b}
N_{ab}
\chi^{\mathrm{I}}_a(\Phi)
\chi^{\mathrm{II}}_b(\Phi)
\label{partition function with flux}
\end{align}
where $\chi^{\mathrm{I,II}}_a(\Phi)$
is a (chiral) partition function
for edge I, II,
and $N_{ab}$ is some coefficient.
Each $\chi_a(\Phi)$ is not invariant under
$\Phi\to \Phi+n\Phi_0$
(``spectral flow''),
while the total partition function should be invariant.
This gauge argument by Laughlin
suggests the stability of the QHE
against disorder and interactions.
In the case of the QSHE, flux insertion argument
can also be applied, and it was shown that a flux
of ${\Phi_0}/2$ pumps fermion number parity and 
lead to spin-charge separation.
\cite{qi2008}

To summarize,
for a chiral edge theory of the QHE, charge is not conserved
under an adiabatic process
to achieve a large gauge transformation,
$\Phi\to \Phi + n\Phi_0$,
signaling pumping of electric charge and thus
detecting the bulk topological insulator.
For later purpose,
this observation can be equivalently rephrased as follows:
if we ``force'' a chiral edge theory to conserve
$N_{\mathrm{I}}$ and $N_{\mathrm{II}}$ separately,
where $N_{\mathrm{I}}$ ($N_{\mathrm{II}}$) is the fermion number at edge I (edge II),
then, the edge partition function $Z(\Phi)$
cannot be made invariant
under $\Phi\to \Phi + \Phi_0$.

\subsection{Large coordinate transformations in gravity}

\subsubsection{Perturbative and global gravitational anomalies}

For systems where electrical charge is not conserved,
we cannot rely on
U(1) gauge (non-) invariance
of the edge theory to diagnose the stability of the topological phase.
A natural tool to address the stability/instability is, then,
(non-) invariance under diffeomorphism transformations
(coordinate transformations).
(See, for example,
Refs.\ \onlinecite{RyuMooreLudwig2012,WnagQiZhang2011}
and references therein).

Similar to the electromagnetic U(1) gauge field in
non-simply connected geometry,
there are infinitesimal as well as large coordinate transformations
when the spacetime manifold
has non-trivial topology.
That is, coordinate transformations that can be reached by
successive infinitesimal transformations from
the identity, and those that are not continuously
connected to the identity, respectively.

The non-invariance of the system under
infinitesimal coordinate transformations
(``perturbative gravitational anomaly'')
means the violation of energy-momentum conservation,
$\langle D^{\mu} T_{\mu\nu}\rangle \neq 0$,
where $T_{\mu\nu}$ is the energy-momentum tensor
and
$D^{\mu}$ is the covariant derivative.
When this happens at the boundary of some bulk system,
the fact that energy-momentum cannot be made conserved within
the boundary theory necessitates the presence of the bulk theory;
energy-momentum at the boundary should be ``leaking'' into the bulk,
and in fact this bulk is what we call a topological phase.
(See, for example,
Refs.\ \onlinecite{RyuMooreLudwig2012,WnagQiZhang2011},
and also Ref.\ \onlinecite{Stone2012}).
For example, the chiral edge theory of a (fractional) quantum Hall fluid
is anomalous under infinitesimal coordinate transformations.
\cite{AlvarezGaumeWitten1983}
This signals the topological property of the bulk
with non-zero thermal Hall conductance $\kappa_{xy}$.
\cite{Volovik90,readgreen,Cappelli01}

Even when there is no perturbative gravitational anomaly,
{\it e.g.}, when the edge theory in question is non-chiral
as in our example of the topological phases with
$\mathbb{Z}_2\times \mathbb{Z}_2$ symmetry,
the system may not be invariant under
{\it large} diffeomorphism transformations
(``global gravitational anomaly''\cite{Witten1985}).
Similarly to perturbative gravitational anomaly,
we will argue below that the non-invariance of the edge theory under
large coordinate transformations
can also be used
as a diagnose of the stability/instability of
the topological phase.

\subsubsection{Modular transformations on a torus}

More specifically,
we again assume the bulk is defined on a finite cylinder with two edges.
The edges may support a chiral or non-chiral edge mode, 
which we assume is a chiral or non-chiral CFT.
The CFT on one edge is defined on a
torus
$T^2=S^1\times S^1$
with
the periodically identified spatial coordinate
(parameterizing the edge),
and
the periodically identified (imaginary) time.

There are a set of large coordinate transformations
on a two-dimensional torus, {\it modular transformations},
which form a group, $\Gamma$.
\cite{Polchinski98}
The geometry of a flat torus is specified by
two real parameters (``moduli''),
or a single complex parameter
$\tau = \omega_2 /\omega_1$,
the ratio of the two periods of the torus
($\mathrm{Im}\,\tau > 0$).
Two different modular parameters
$\tau$ and $\tau'$
can describe the
same toroidal geometry
if they are related by
an integer linear transformation with
unit determinant,
\begin{align}
\tau' = \frac{ a \tau+b}{c\tau + d},
\quad
a,b,c,d\in \mathbb{Z},
\quad
ad-bc=1.
\end{align}
(Here, $\tau$ should not be confused with
the imaginary time).
Modular transformations belong
to the infinite discrete group
$\mathrm{PSL}(2, \mathbb{Z}) =
\mathrm{SL}(2, \mathbb{Z})/\mathbb{Z}_2$.
These transformation
are generated by two
generators,
$T : \tau \to \tau +1$ and
$S : \tau \to  −1/\tau$ ,
satisfying the relations
$S^2 = (ST )^3 = C$,
where $C$ is the charge conjugation matrix, satisfying
$C^2 = 1$.

For a CFT on a torus,
we can ask if it is invariant under modular transformations.
Any CFT that is derived from the continuum limit
of a two-dimensional lattice statistical mechanical system
(or equivalently a one-dimensional lattice quantum system)
is expected to be
anomaly free (modular invariant).
\cite{Cardy1986}
On the contrary,
if a CFT in question is not modular invariant,
it may not be realized, on its own,
as a continuum limit of a local lattice system, 
and
must be accompanied by some (topological) bulk theory.

Based on these observations,
we are lead to claim that
the global gravitational anomaly 
implies the presence of a topological bulk theory,
in a way quite analogous to 
the previous illustration of the charge response;
basically, we simply replace
$\Phi$ by $\tau$,
and the large gauge transformation
$\Phi \to \Phi+\Phi_0$
by modular transformations,
$\tau\to \tau+1$ and $\tau\to -1/\tau$.
The partition function now depends on a complex parameter
$\tau$ (the moduli parameter of the torus),
$Z(\tau,\bar{\tau})$.
The modular non-invariance of the partition function of
a given edge signals the presence of a topological bulk theory.
Note, however, that
when the two edges (edge I and edge II) are combined,
we should be able to achieve
the modular invariance;
\cite{Cappelli96,Cappelli09,Cappelli09b}
they can be gapped pairwise.
Similarly to
Eq.\ (\ref{partition function with flux}),
we can write the total partition function
in terms of a liner combination
\begin{align}
Z (\tau,\bar{\tau})
=
\sum_{a, b}
N_{ab}
\chi^{\mathrm{I}}_a(\tau,\bar{\tau})
\chi^{\mathrm{II}}_b(\tau,\bar{\tau}).
\end{align}
Each block 
$\chi^{\mathrm{I},\mathrm{II}}_a(\tau,\bar{\tau})$
can be non-modular invariant, but the total
partition function 
$Z(\tau,\bar{\tau})$
should be modular invariant.

If the system is defined, 
at the microscopic level, 
in terms of fermions (electrons), 
the requirement that the total partition function 
$Z(\tau,\bar{\tau})$
to be modular invariant 
may be relaxed; 
In the presence of fermions, 
the partition function 
may not be invariant under $T$, but 
should still be invariant under $T^2$.
\cite{Cappelli96,Cappelli09,Cappelli09b}
The modular transformations
generated by $S$ and $T^2$ 
form a subgroup [$=\Gamma(2)$] of the full modular 
group $\Gamma$.

\subsubsection{Symmetry projection}

When there is a set of symmetries,
and when we talk about symmetry-protected topological phases,
it makes sense to diagnose the system by
an adiabatic process which does not violate
the symmetries.

For a unitary symmetry,
a convenient way to enforce the symmetry
in the adiabatic process is to
{\it project} the total Hilbert space
into a given subsector specified by a
quantum number.
We then ask if, for a given edge separately,
each sector can be made modular invariant
({\it i.e.}, free of global gravitational anomaly).

Inability to achieve this would mean
a quantum number of some kind should
be ``pumped'' from one edge to the other
along an adiabatic process to generate
a modular transformation;
When both edges are included,
the total systems without projection would be modular invariant.
This would mean
the symmetry (conservation of a quantum number)
should be violated
in the adiabatic process, and thus leads to pumping.

Let us have a further look at the projection procedure.
When projected,
certain states
(states which are not singlet under a symmetry group in question)
are removed from the original Hilbert space
of the edge theory.
From the state-operator correspondence in CFT,
this means the corresponding operators
are not allowed in the theory after projection.
Such operators,
$\mathcal{O}(z,\bar{z})$, say,
in the original theory, can be added to
the action $S_0$ describing the edge theory
as a perturbation,
$S_0\to S_0 + \lambda \int d^2x\,\mathcal{O}(z,\bar{z})$,
where $\lambda$ is a coupling constant,
and if $\mathcal{O}(z,\bar{z})$ is relevant
in the renormalization group (RG) sense, it can destabilize the edge.
As its corresponding state,
the operator is not singlet under the symmetry group,
and hence when added to the action,
it explicitly breaks the symmetry.
In the projected theory,
such perturbations are prohibited.

\subsection{Free complex fermion}
\label{free complex fermion}

To illustrate 
the spectral flow 
(non-invariance under large gauge transformations)
and
the modular non-invariance
(global gravitational anomaly),
and also for our later use,
let us consider
a single copy of
left-moving {\it complex} fermion
as an example.
(We follow Refs.\ \onlinecite{Polchinski98,Ginsparg91,Imamura2011}.)
It is described by the Lagrangian
\begin{align}
\mathcal{L}_L
 =
 \frac{1}{2\pi}
 \Psi^{\dag}_L
\left( \partial_{\tau} + v \partial_x \right)
\Psi^{\ }_L.
\label{lagrangian complex fermion}
\end{align}
The path integral for a single copy of complex fermion
can be considered with boundary conditions in space and time
directions:
\begin{align}
\Psi_L(\tau, x+\ell) &= (-1)e^{ 2\pi {i}\lambda} \Psi_L(\tau, x), 
\nonumber \\
\Psi_L(\tau+T^{-1},x) &= (-1)e^{ -2\pi {i}\mu} \Psi_L(\tau, x), 
\end{align}
where $T^{-1}$ is the inverse temperature, 
and the system is defined on a spatial circle of circumference 
$\ell$;
$\mu$ and $\lambda$ specify the boundary condition 
for the space and time directions, respectively. 
In particular, if the chiral Lagrangian
(\ref{lagrangian complex fermion})
is interpreted as the edge theory of the IQHE, 
$\lambda$ is related to the flux $\Phi$
in 
Sec.\ \ref{large gauge transformations in electromagnetism}
as
$\Phi/\Phi_0 = \lambda$. 
The corresponding partition function is denoted as 
\begin{align}
Z^{\lambda}{ }_{\mu}(\tau).
\end{align}
Here, 
$\tau$ is the modular parameter,
and 
the upper script indicates the boundary condition
in space direction whereas the lower script
indicates the boundary condition
in time direction.
Later, when we consider real (Majorana) fermions, rather than complex fermions, 
we also use notation
``0 (1/2)''= ``A (P)''
= antiperiodic (periodic) boundary condition.

The fermionic path integral 
(fermionic determinant) 
is evaluated as 
\begin{align}
& Z^{\lambda}{ }_{\mu}(\tau)
=
e^{2\pi {i}\lambda \mu}
q^{-1/24}
q^{\lambda^2/2}
\nonumber \\
&\qquad
\times 
\prod^{\infty}_{n=1}
\left( 1 +
w
q^{n-1/2}
\right)
\left( 1 +
w^{-1}
q^{n-1/2}\right), 
\label{eq: fermionic partition function}  
\end{align}
where
$
w = e^{2\pi{i} \mu}q^{\lambda}
$.
Here, the overall phase factor
$e^{2\pi {i}\lambda \mu}$ is purely 
conventional;
Since we have 
an independent path integral for 
a given set of boundary conditions,
there is no unique way to 
determine the relative (Boltzmann) weight 
between sectors with different boundary conditions. 
The factor $e^{2\pi {i}\lambda \mu}$
in Eq.\ (\ref{eq: fermionic partition function})  
is a common choice, but this will not 
affect out discussion below. 

\subsubsection{Spectral flow}

Let us derive 
Eq.\ (\ref{eq: fermionic partition function})
in the operator formalism,
where
the partition function with 
given boundary conditions is given by
\begin{eqnarray}
Z^{\lambda}{ }_{\mu}(\tau)
=
\mathrm{Tr}_{\lambda}
\left[
e^{-2 \pi {i} \mu N_L}
q^{H_L}
\right],
\quad
q= e^{2\pi {i} \tau},
\end{eqnarray}
where
$\mathrm{Tr}_\lambda$
is the trace over the Hilbert space 
defined with the spatial boundary condition $\lambda$.
Here,
\begin{align}
N_L :=
\int^{\ell}_0 dx\,  \Psi^{\dag}_L \Psi^{\ }_L
\end{align}
is the total left-moving fermion number.
Observe that, in the operator formalism,
the periodic boundary condition in time is realized here by
an insertion of operator
$e^{ -2 \pi {i} \mu N_L}$.

The partition function can be evaluated explicitly by 
making use of the mode expansion
\begin{align}
\Psi_L(x)&=
\sqrt{\frac{2\pi}{\ell}}
\sum_{s\in \mathbb{Z}+1/2 -\lambda} 
e^{-{i}x \frac{2\pi s}{\ell}}
\Psi_{s}^{\ },
\end{align}
where $\Psi_{s}^{\ }$ and $\Psi_{s}^{\dag}$ satisfy 
the commutation relation
$\{
\Psi_{s}^{\ },\Psi_{s'}^{\dag}
\}=\delta_{ss^{\prime}}
$. 
In terms of the mode expansion,
we define the ground state 
$\left.|0\right>_{\lambda}$
for a given spatial boundary condition $\lambda$ 
as 
a filled Dirac sea, 
\begin{align}
&
\Psi_{n+1/2-\lambda}
\left.|0\right>_{\lambda}=
\Psi^{\dag}_{-n-1/2+\lambda}
\left.|0\right>_{\lambda}=0
\nonumber \\
&\quad 
\mbox{for}
\quad 
n+1/2-\lambda> 0.
\end{align}
Let us further assign the fermion number to the ground state
as
\begin{align}
e^{ {i} \phi N_L}
|0\rangle_{\lambda}
=
e^{ {i} \phi  \lambda}
|0\rangle_{\lambda},
\quad 
\phi \in \mathbb{Z}.
\end{align}
Similarly to discussion below
Eq.\ (\ref{eq: fermionic partition function}), 
this assignment is purely conventional. 
With this assignment, we obtain
the partition function 
(\ref{eq: fermionic partition function}).

The two boundary conditions
$\lambda_1$
and
$\lambda_2$
are in general physically distinct,
and correspondingly, 
the two ground states,
$|0\rangle_{\lambda_1}$
and
$|0\rangle_{\lambda_2}$,
belong to different Hilbert spaces.
However, 
when $\lambda_1 -\lambda_2=
\mbox{(integer)}$,
these two systems are related by 
a large gauge transformation. 
Let us now consider 
an adiabatic process interpolating two boundary conditions, 
$\lambda=0\to \lambda=1$, say. 
While these boundary conditions are large-gauge equivalent, 
the ground state might not evolve into 
itself (the ground state) under the adiabatic process
(``spectral flow''): 
For example, let us start from
$\lambda=0$ and define the ground state as
\begin{align}
&
\Psi_{n+1/2}
\left.|0\right>_{\lambda=0}=
\Psi^{\dag}_{-n-1/2}
\left.|0\right>_{\lambda=0}=0
\nonumber \\
&
\quad
\mbox{for}
\quad 
n \ge 0.
\end{align}
As we change $\lambda$, 
we assume the ground state evolves continuously:
it is always annihilated by 
$\Psi_{n+1/2-\lambda}$ with $n\ge 0$. 
We define the state obtained by this 
adiabatic process as
$| 0'\rangle_{\lambda}$
On the other hand, by definition,
the ground state at $\lambda=1$ is given by 
\begin{align}
&
\Psi_{n-1/2}
\left.|0\right>_{\lambda=1}=
\Psi^{\dag}_{-n+1/2}
\left.|0\right>_{\lambda=1}=0
\nonumber \\
&
\quad
\mbox{for}
\quad 
n\ge 1,
\end{align}
{\it i.e.}, 
it is annihilated by 
$\Psi_{n+1/2-\lambda}$ with $n=1$. 
We conclude 
$|0'\rangle_{\lambda=1}
=
\Psi^{\dag}_{-1/2}
|0\rangle_{\lambda=1}
\neq |0\rangle_{\lambda=1}
$.
This spectral flow is reflected in the non-invariance 
of the partition function
under the adiabatic process.

\subsubsection{modular transformation}

Let us now examine 
the transformation properties of the partition function
under modular transformations:
From Eq.\ (\ref{eq: fermionic partition function}),
\begin{align}
Z^0{ }_{0}(\tau+1)
&=
e^{-{i}\pi/12}
Z^0{ }_{1/2}(\tau),
\nonumber \\
Z^0{ }_{1/2}(\tau+1)
&=
e^{-{i}\pi/12}
Z^0{ }_{0}(\tau),
\nonumber \\
Z^{1/2}{ }_{0}(\tau+1)
&=
e^{{i}\pi/6}
Z^{1/2}{ }_{0}(\tau),
\nonumber \\
Z^{1/2}{ }_{1/2}(\tau+1)
&=
e^{{i}\pi/6}
Z^{1/2}{ }_{1/2}(\tau), 
\\
\nonumber \\
Z^{0}{ }_{0}(-1/\tau)
&=
Z^{0}{ }_{0}(\tau),
\nonumber \\
Z^{0}{ }_{1/2}(-1/\tau)
&=
Z^{1/2}{ }_{0}(\tau),
\nonumber \\
Z^{1/2}{ }_{0}(-1/\tau)
&=
Z^{0}{ }_{1/2}(\tau),
\nonumber \\
Z^{1/2}{ }_{1/2}(-1/\tau)
&=
e^{ -\pi {i}/2}
Z^{1/2}{ }_{1/2}(\tau).
\end{align}
The partition function
$Z^{1/2}{ }_{1/2}(\tau)$ is actually zero identically,
because of the zero mode of the Dirac operator with
periodic boundary condition in both directions.
Nevertheless, we have assigned formal transformation
rules to $Z^{1/2}{ }_{1/2}(\tau)$.

The transformation law for
$
\tau \to -1/\tau
$
is what we expect classically
({\it i.e.}, just exchanging space and time boundary conditions),
but the transformation law for
$
\tau \to \tau + 1
$
is somewhat unexpected in the sense that
the partition function acquires a phase factor.
The reason for this is that there is
no diff-invariant way to define the phase of
the path integral for purely left-moving
fermions. For left- plus right-moving
fermions with matching boundary conditions,
the path integral can be defined by
Pauli-Villars or other regulators.
This is the same as the absolute square of the
left-moving path integral, but leaves
a potential phase ambiguity
in that path integral separately.
The phase represents a global gravitational anomaly,
an inability to define the phase of the path integral
such that it is invariant under large
coordinate transformations.
Of course, a single-left moving fermion
has non-zero chiral central charge
and so has an anomaly even under infinitesimal coordinate
transformations, but the global
anomaly remains even when a left- and right-moving
fermion are combined (see below).

\section{Edge theory
of $\mathbb{Z}_2\times \mathbb{Z}_2$
symmetric topological phase}
\label{edge theory of Z2xZ2 symmetric topological phase}

Let us now consider the edge theory of
the $\mathbb{Z}_2\times \mathbb{Z}_2$ symmetric topological phase,
Eq.\ (\ref{edge theory N=8}). 
We focus on the case of $N_f=2N$
and demonstrate that while when $N\neq 4$ (mod 4),
there is a global gravitational anomaly,
the case with $N=4$ (mod 4) is anomaly free.
In fact, this is deeply related to the modular invariance
and the consistency of type II superstring theory.
\cite{Imamura2011}
[While our presentation below uses,
in order to make use of our discussion in 
Sec.\ \ref{free complex fermion},
the partition function $Z^{\lambda}{ }_{\mu}(\tau)$
of a complex fermion,
there is no fundamental reason to do so.
The entire discussion can be constructed in terms of
real (Majorana) fermions, without referring to complex fermions.]

Since there are various boundary conditions allowed for
the fermionic edge theory,
the partition function is given
as a sum of sectors with different boundary conditions.
Let us discuss this issue by using the operator formalism.
By considering contributions from different spatial boundary
conditions, we consider a sum
\begin{align}
\sum_{\alpha} \mathrm{Tr}_{\alpha}\,
\left[q^{H_{\alpha}}
\right]
\label{sum space}
\end{align}
where the summation extends all possible spatial boundary conditions,
and $H_{\alpha}$ is the Hamiltonian with a boundary condition specified 
by $\alpha$.
(Here in our problem, $\alpha=\mathrm{A}, \mathrm{P}$).
Since the modular transformation exchanges the spatial and time
directions, Eq.\ (\ref{sum space}) is not modular invariant;
we have to consider contributions from
different boundary conditions in the time direction as well.
As we have seen, in the operator
formalism, a different kind of boundary in time direction is
achieved by an insertion of a unitary operator.
Thus, the partition function is given by
\begin{align}
Z=
\sum_{\alpha,\alpha'} \mathrm{Tr}_\alpha\,
\left[U_{\alpha'}
q^{ H_{\alpha}}
\right]
\label{partition function, all bc}
\end{align}
where $U_{\alpha}$ is some unitary operator.
(In our case, $U_{\alpha}$ is the parity of
the fermion number operators.)
The partition function can also be written as
\begin{align}
&
\quad
Z
=
\mathcal{N}
\sum_{\alpha} \mathrm{Tr}_{\alpha}\,
\left[
P q^{H_{\alpha}}
\right],
\nonumber \\
&\mbox{where}
\quad
P :=
\frac{1}{\mathcal{N}} \sum_{\alpha} U_\alpha.
\end{align}
Under the assumption that the set of
unitary operators $\{U_{\alpha} \}_{\alpha=1,\ldots, \mathcal{N}}$ form a group,
one verifies that
\begin{align}
U_\alpha P = P U_\alpha = P^2 =P.
\end{align}
Thus, $P$ is a projection operator.

As we have seen, for the fermionic edge theory,
the unitary operators that we need to change boundary conditions
are the fermion number parity operators,
\begin{align}
U_{\alpha}
=
1,
\quad
(-1)^{N_L},
\quad
(-1)^{N_R},
\quad
(-1)^{N_L+N_R},
\end{align}
where
$N_L =\sum_{i=1}^N N^i_{L}$
and
$N_R =\sum_{i=1}^N N^i_{R}$
are the total
left- and right-moving fermion number,
respectively.
The sum (the projection operator) is then
\begin{align}
P
&=
\frac{1}{4}
\left[
1+
(-1)^{N_L}
+
(-1)^{N_R}
+
(-1)^{N_L+N_R}
\right]
\nonumber \\
&=
\frac{1+(-1)^{N_L}}{2}
\times
\frac{1+(-1)^{N_R}}{2}
\nonumber \\
&=:P^{\ }_{\mathrm{GSO}}.
\label{GSO}
\end{align}
This operator projects, for each of the left- and right-moving sectors,
onto the space of a definite fermion number parity
[the Gliozzi-Scherk-Olive (GSO) projection].
Observe that this projection
acts on the left- and right-moving sectors
separately.

For the left-moving sector with
$\alpha=0=\mathrm{A}$ 
spatial boundary condition
in Eq.\ (\ref{partition function, all bc}),
\begin{align}
Z_{\mathrm{A}}(\tau)
&=
\mathrm{Tr}_{\mathrm{A}}\,
\left[
P^{\ }_{\mathrm{GSO}} q^{H_{\mathrm{A}}}
\right]
=
\mathrm{Tr}_{\mathrm{A}}
\big[
P^{\ }_{\mathrm{GSO}}
q^{H^1_{\mathrm{A}}+\cdots + H^N_{\mathrm{A}}}
\big]
\nonumber \\
&=
\frac{1}{2}
\mathrm{Tr}_{\mathrm{A}}
\big[
q^{H^1_{\mathrm{A}}+\cdots + H^N_{\mathrm{A}}}
\big]
\nonumber \\
&\quad 
+
\frac{1}{2}
\mathrm{Tr}_{\mathrm{A}}
\big[
e^{\pi {i} N_L} q^{H^1_{\mathrm{A}}+\cdots + H^N_{\mathrm{A}}}
\big]
\nonumber \\
&=
\frac{1}{2}
\left[
Z^{0}{ }_0(\tau)^N
\pm
Z^{0}{ }_{1/2}(\tau)^N
\right].
\end{align}
The sign $\pm$  in the last line 
indicates a possible ambiguity
in assigning the fermion number parity 
to the ground state $|0\rangle_{\mathrm{A}}$
in the $\alpha=\mathrm{A}$ sector;
see discussion around Eq.\ 
(\ref{eq: fermionic partition function}).
While we adopted a particular choice for 
the fermion number parity in
Eq.\ (\ref{eq: fermionic partition function}),
here we leave other possibilities open 
in order to illustrate
such ambiguity does not affect our conclusion.
Similarly, for
$\alpha=1/2=\mathrm{P}$ 
spatial boundary condition,
\begin{align}
Z_{\mathrm{P}}(\tau)
&=
\mathrm{Tr}_{\mathrm{P}}\,
\left[
P^{\ }_{\mathrm{GSO}} q^{H_{\mathrm{P}}}
\right]
\nonumber \\
&=
\frac{1}{2}
\big[
Z^{1/2}{ }_0(\tau)^N
\pm
Z^{1/2}{ }_{1/2}(\tau)^N
\big].
\end{align}
There is again a sign ambiguity $\pm$ here,
regarding to the fermion number parity 
of the ground state in the 
$\alpha=\mathrm{P}$ sector.

The total partition function for the $N_f=2N$ left-moving
Majorana fermions 
$Z_L(\tau)$ is obtained by 
taking a linear combination of
$Z_{\mathrm{A}}(\tau)$
and
$Z_{\mathrm{P}}(\tau)$.
The requirement
that the total partition function
is invariannt under the $S$-modular transformation 
motivates us to consider 
the following relative weight between
$Z_{\mathrm{A}}(\tau)$
and
$Z_{\mathrm{P}}(\tau)$:
\begin{align}
Z_L(\tau)
&=
\frac{1}{2}
\big[
Z^{0}{ }_0(\tau)^N
+ s
Z^{0}{ }_{1/2}(\tau)^N
\nonumber \\
&\quad 
+ s
Z^{1/2}{ }_0(\tau)^N
+ s s'
Z^{1/2}{ }_{1/2}(\tau)^N
\big],
\end{align}
where the signs
$s,s'=\pm 1$
are related to the ambiguity of the 
fermion number parity of the 
ground states $|0\rangle_{\mathrm{A},\mathrm{P}}$,
and to the relative weight between
$Z_{\mathrm{A}}(\tau)$
and
$Z_{\mathrm{P}}(\tau)$
when taking a linear combination.

Under $T$-modular transformation, 
the partition function is transformed as
\begin{align}
&
Z_L(\tau)
=
s 
e^{{i} \frac{\pi N}{12}}
\frac{1}{2}
\left[
(Z^{0}{ }_0)^N
+ s
(Z^{0}{ }_{1/2})^N
\right.
\nonumber \\
&\quad 
\left.
+
e^{-{i} \frac{\pi N}{4}}
(Z^{1/2}{ }_0)^N
+ s' 
e^{-{i} \frac{\pi N}{4}}
(Z^{1/2}{ }_{1/2})^N
\right]
(\tau+1),
\end{align}
whereas under $T^2$, 
\begin{align}
&
Z_L(\tau)
=
e^{{i} \frac{\pi N}{6}}
\frac{1}{2}
\left[
(Z^{0}{ }_0)^N
+s
(Z^{0}{ }_{1/2})^N 
\right.
\nonumber \\
&\quad 
\left.
+s
e^{-{i} \frac{\pi N}{2}}
(Z^{1/2}{ }_0)^N
+ ss' 
e^{-{i} \frac{\pi N}{2}}
(Z^{1/2}{ }_{1/2})^N
\right]
(\tau+2).
\end{align}
Thus, when $N=4$,
we thus achieve the modular covariance,
$Z_L(\tau)
\to
Z_L(\tau)
=
e^{{i}2\pi /3}
Z_L(\tau+2).
$
Combined with the right-moving part of the partition function,
$Z_{R}(\bar{\tau})$, 
the total partition function
$Z(\tau,\bar{\tau})
=
Z_R(\bar{\tau})
Z_L(\tau)
=
|Z_L(\tau)|^2
$
is then
invariant under $T^2$,
\begin{align}
Z(\tau,\bar{\tau})
&=
Z(\tau+2,
\bar{\tau}+2).
\end{align}
Similarly,
when $N=4$,
by choosing $s=-1$, 
we thus achieve the modular covariance,
$Z_L(\tau)
\to
Z_L(\tau)
=
(-1) e^{{i}\pi /3}
Z_L(\tau+1).
$
Combined with the right-moving part of the partition function,
$Z_{R}(\bar{\tau})$, the total partition 
function is then
modular invariant,
\cite{footnote1}
\begin{align}
Z(\tau,\bar{\tau})
&=
Z(\tau+1,
\bar{\tau}+1).
\end{align}

In the Lagrangian (\ref{edge theory N=8}),
the fermions $\psi^{a}_{R,L}$ are in the vector representation
of SO(8), $\textbf{8}_v$. In the context of superstring theory,
this is the 
Ramond-Neveu-Schwarz (RNS) model of the superstring
in the light-cone gauge.
The Lagrangian does not completely 
specify
the spectrum, and we need to impose the boundary conditions;
the fermions $\psi^{a}_{R,L}$ obey either
antiperiodic (NS) or periodic (R) boundary condition.
Furthermore, we have used the GSO projection (\ref{GSO}),
which leads to type IIB and type IIA theories.
Because of triality, 
one can rewrite the 
$\psi^{a}_{R,L}$ theory in
terms of spinors $\xi^{a}_{R,L}$ and $\eta^{a}_{R,L}$ as well.
Technically, this means we first bosonize
the RNS fermions $\psi^{a}_{R,L}$, and refermionize,
to obtain $\xi^a$ and $\eta^a$,
spinor ($\textbf{8}_s$)
and 
conjugate spinors ($\textbf{8}_c$)
-- this is the Green-Schwarz (GS) formalism of the superstring.
The two spinors,
$\xi^a$ and $\eta^a$
are distinguished by chirality operator
of SO(8);
spinor $\xi^a$ has positive chirality
and
conjugate spinor $\eta^a$ has negative chirality.
When, rewritten in terms of these spinors,
in type IIB theory,
we have left-moving and right-moving spinors,
and the Lagrangian is given by
\begin{align}
\mathcal{L}
&=
\frac{1}{4\pi}
\sum^{N_f=8}_{a=1}
\big[
\xi^a_L (\partial_{\tau} + {i}v \partial_x) \xi^a_L
+
\xi^a_R (\partial_{\tau} - {i}v \partial_x) \xi^a_R
\big].
\end{align}
Similarly, in type IIA theory,
we have
left-moving spinor and
right-moving conjugate spinors,
and the Lagrangian is given by
\begin{align}
\mathcal{L}
&=
\frac{1}{4\pi}
\sum^{N_f=8}_{a=1}
\big[
\xi^a_L  (\partial_{\tau} +{i}v \partial_x) \xi^a_L
+
\eta^a_R (\partial_{\tau} -{i} v \partial_x) \eta^a_R
\big].
\end{align}
Unlike the vector fermions $\psi^{a}_{R,L}$,
the spinors obey periodic boundary condition only:
\begin{align}
\xi^a_L(x+\ell) =
\xi^a_L(x),
\quad
\xi^a_R(x+\ell) =
\xi^a_R(x),
\nonumber \\
\eta^a_L(x+\ell) =
\eta^a_L(x),
\quad
\eta^a_R(x+\ell) =
\eta^a_R(x),
\end{align}
where the system is defined on a spatial circle of circumference
$\ell$.
Because of this, there is no need for projection.
One can compare the spectrum of
the
RNS theory with GSO projection,
and
the GS theories;
they match precisely.

We conclude this section with a discussion on the ``Ising projection".
As emphasized before,
we have two separate projections for the left- and right-moving sectors.
This should be contrasted to the projection
with respect to the total fermion parity
$(-1)^{N_L +N_R}$
which is described by the ``diagonal'' projection operator
\begin{align}
P_{0}
=
\frac{1 + (-1)^{N_L +N_R}}{2}.
\end{align}
For $2N$ flavor of Majorana fermions,
the resulting total partition function
\begin{align}
&
\frac{1}{2}
\big\{
|Z^{0}{ }_{0}(\tau)|^{N}
+
|Z^{0}{ }_{1/2}(\tau)|^{N}
\nonumber \\
&\quad 
+
|Z^{1/2}{ }_{0}(\tau)|^{N}
\mp
|Z^{1/2}{ }_{1/2}(\tau)|^{N}
\big\}
\end{align}
is invariant for any $N$
because the phases cancel in the absolute values.
The Ising model can be viewed as an example of
the above
partition function. 
(Only minor difference is that
we have been mainly using the complex fermions,
instead of Majorana fermions.)
The Ising partition function is given by
\begin{align}
Z_{\mathrm{Ising}}
&=
\frac{1}{2}
\left\{
|\chi^0 { }_0|^2
+
|\chi^{1/2} { }_0|^2
+
|\chi^0 { }_{1/2}|^2
\pm
|\chi^{1/2} { }_{1/2}|^2
\right\}. 
\end{align}
Here, $\chi^{\lambda}{ }_{\mu}(\tau)$
is the partition function of
a left-moving Majorana (not complex) fermion
with boundary conditions specified by $\lambda$
and $\mu$.
As illustrated above,
this partition function can be obtained by
considering the following projection: 
$
Z_{\mathrm{Ising}}
=
\mathrm{Tr}_{\mathrm{A}\oplus \mathrm{P}}\,
\left[
P_0 \, 
q^{H_L}\bar{q}^{H_R}
\right]
$.
\cite{footnoteIsing}

\section{Discussion}

The modular invariance plays a major role in CFT
\cite{Cardy1986,CappelliItzyksonZuber,Kato1987}
and also in string theory.
Its importance in chiral topological phases such as
the fractional QHE has also been emphasized.
\cite{Cappelli09,Cappelli09b}

Partly motivated by recent discoveries of non-chiral topological phases, 
\cite{Nepert2011}
such as the QSHE,
we studied in this paper
an implication of modular invariance in
non-chiral topological phases protected by discrete symmetries.
Quite generically, a non-chiral edge theory can be gapped by
some perturbation
by ``coupling'' the left- and right-moving sectors.
This is implied from the fact that a non-chiral CFT,
when its left- and right-moving parts are properly combined,
can be made modular invariant.
In the presence of a certain symmetry condition,
however, there is a constraint on perturbations
which are allowed to be added to the action.
In an extreme case, the symmetry constraint completely
removes perturbations, in which case
the gapless nature of the edge theory can be protected.
This suggests that if the way we glue
the left- and right-moving 
sectors were to be consistent with
the symmetry condition, we would not be able to achieve
modular invariance.
For the particular example we investigated in this work, 
there is $\mathbb{Z}_2\times\mathbb{Z}_2$ symmetry 
which allows us to decompose the Hilbert space into
different sectors with different quantum numbers.
After this decomposition, we studied if each sector
can be made modular invariant separately.
Even though we have looked at a particular example
of the $\mathbb{Z}_2\times\mathbb{Z}_2$ symmetric topological phase,
we expect the proposal using the modular invariance as a
diagnostic tool for more general topological phases
without local 
(perturbative) anomalies.

We close with several comments. 
(i)
For the bulk of the paper, we have discussed
mainly modular invariance/non-invariance
of non-chiral CFTs.
A chiral CFT can also be
modular invariant/non-invariant on its own as well.
A well-known example is
a collection of
$N$ copies of chiral complex fermions
or
$2N$ copies of chiral Majorana fermions.
Let us consider the partition function
given by the following combination:
\begin{align}
&\quad 
\frac{1}{2}
\left\{
\big[Z^{0}{ }_{0}(\tau)\big]^{N}
+
\big[Z^{0}{ }_{1/2}(\tau)\big]^{N}
+
\big[Z^{1/2}{ }_{0}(\tau)\big]^{N}
\right\}
\nonumber \\
&=
\frac{1}{2}
\left\{
e^{ {i}N\pi/12}
\big[Z^{0}{ }_{1/2}\big]^{N}
+
e^{ {i}N\pi/12}
\big[Z^{0}{ }_{0}\big]^{N}
\right.
\nonumber \\
&
\left.
\qquad
\qquad 
+
e^{ -{i}N\pi/6}
\big[Z^{1/2}{ }_{0}\big]^{N}
\right\}
(\tau+1).
\end{align}
The chiral central charge is $c_L=N$.
The partition function is clearly
$S$-modular invariant.
In order to achieve
invariance under $T$-transformation,
we
need,
at least,
$
N = 8k
$
copies of fermions,
where $k$ is a positive integer.
If we consider
$16k$ chiral Majorana fermions
or
$8k$ complex fermions,
the partition function is modular {\it covariant}.
In particular, when $k=1$,
the chiral central charge is $c_L=8$.
(When bosonized, this is the partition function
of the compactifed bosons on the root lattice $E_8$).
If we cube this partition function,
we achieve the true modular invariance with $c_L=24$.
The chiral topological phase
with $2N$ copies of chiral Majorana fermions
at its edge was discussed in
the context of the honeycomb lattice Kitaev model.
\cite{Kitaev2005}
A similar kind of mod 16 periodicity was
observed in the bulk topological properties
(non-Abelian statistics of quasiparticles in the bulk
depends on the bulk Chern number mod 16).

(ii)
We have used
symmetry projection as a diagnostic
tool to study the stability of
non-interacting, 
symmetry-protected, topological phases.
Instead, it is also possible to think of a
topological phase with gauge interactions
in the bulk.
In this case,
projections are performed dynamically 
in the bulk and in the edge theories.
One of such models in the bulk would 
look like the two copies of
the honeycomb lattice Kitaev model
\cite{Kitaev2005}
with opposite chiralities.

(iii)
While robust in the presence of a certain set of symmetries, 
non-chiral edges are in general susceptible to symmetry breaking
perturbations. 
In particular, one can study the response of the edge theory to 
a local perturbation, such as a single impurity, 
or to a topological defect at the edge,
which would reflect topological properties of the bulk.
(See, for example, Refs.\ \onlinecite{QiHughesZhang2007,Maciejko2009}
for the edge state of the QSHE.) 
For the $\mathbb{Z}_2\times \mathbb{Z}_2$
symmetric topological phase,
such local impurity problems in the edge state,
in the long-wave length limit, may correspond
to D-branes.

(iv)
Finally, 
there are topological phases 
that are not accompanied 
by a gapless edge state. 
Whether or not these topological phases
can be understood in terms of quantum anomalies of some kind
is an open question.

\acknowledgements

We would like to thank Xiao-Liang Qi
for sharing his results with us prior to arXiv submission.
We also thank Hong Yao
for fruitful collaboration in a closely related project.
Useful comments from Rob Leigh and Tadashi Takayanagi
are also greatly acknowledged. 
SCZ is supported by the NSF under grant No.\ DMR-0904264.




\begin{thebibliography}{99}

\bibitem{review_QHE}
{\it The Quantum Hall Effect},
edited by R. E. Prange and S. M. Girvin (Springer, New York, 1987).

\bibitem{review_TIa}
M. Z. Hasan, and C. L. Kane,
Rev. Mod. Phys. \textbf{82}, 3045 (2010).

\bibitem{review_TIb}
X.-L. Qi, and S.-C. Zhang,
Rev. Mod. Phys. \textbf{83}, 1057 (2011).


\bibitem{KaneMele}
C.\ L.\ Kane and E.\ J.\ Mele,
Phys.\ Rev.\ Lett. \textbf{95}, 146802 (2005);
Phys.\ Rev.\ Lett. \textbf{95}, 226801 (2005).

\bibitem{Bernevig05}
B.\ A.\ Bernevig and S.-C. Zhang,
Phys.\ Rev.\ Lett. \textbf{96}, 106802 (2006).

\bibitem{Bernevig06}
B.\ A.\ Bernevig, T. Hughes and S.-C. Zhang,
Science \textbf{314}, 1757 (2006).

\bibitem{Moore06}
J.\ E.\ Moore and L.\ Balents,
Phys.\ Rev.\ B \textbf{75}, 121306(R) (2007).

\bibitem{Roy3d}
R.\ Roy,
Phys.\ Rev.\ B \textbf{79}, 195322 (2009).

\bibitem{Fu06_3Da}
L.\ Fu, C.\ L.\ Kane, and E.\ J.\ Mele,
Phys.\ Rev.\ Lett. \textbf{98}, 106803 (2007).

\bibitem{Fu06_3Db}
L.\ Fu and C.\ L.\ Kane,
Phys.\ Rev.\ B \textbf{76}, 045302 (2007).

\bibitem{qilong}
X.-L. Qi, T. L. Hughes and S.-C. Zhang,
Phys. Rev. B
\textbf{78},
195424,
(2008).

\bibitem{Schnyder2008}
A. P. Schnyder, S. Ryu, A. Furusaki, and A. W. W. Ludwig,
Phys. Rev. B \textbf{78}, 195125 (2008).

\bibitem{SRFLnewJphys}
S. Ryu, A. Schnyder, A. Furusaki and A. W. W. Ludwig,
New J. Phys.
\textbf{12},
065010
(2010).

\bibitem{Kitaev2009}
A.\ Yu Kitaev,
\textit{AIP Conf.\ Proc.} \textbf{1134}, 22 (2009).

\bibitem{Wang2010}
Zhong Wang, Xiao-Liang Qi and Shou-Cheng Zhang,
Phys.\ Rev.\ Lett. \textbf{105}, 256803 (2010).

\bibitem{FidkowskiKitaev2009}
L.\ Fidkowski and A.\ Kitaev,
Phys.\ Rev.\ B \textbf{81}, 134509 (2010).

\bibitem{FidkowskiKitaev2010}
L.\ Fidkowski and A.\ Kitaev,
Phys.\ Rev.\ B \textbf{83}, 075103 (2011).

\bibitem{TurnerPollmannBerg2010}
A.\ M.\ Turner, F.\ Pollmann and E.\ Berg,
Phys.\ Rev.\ B \textbf{83}, 075102 (2011).

\bibitem{Gu12}
Zheng-Cheng Gu, and Xiao-Gang Wen, 
\texttt{arXiv:1201.2648}.

\bibitem{Qi12}
Xiao-Liang Qi,
\texttt{arXiv:1202.3983}. 


\bibitem{YaoRyu12}
Hong Yao
and
Shinsei Ryu,
\texttt{arXiv:1202.5805}. 

\bibitem{GuLevin}
Zheng-Cheng Gu and Michael Levin,
unpublished.



\bibitem{Wu2006}
See, for the effects of interactions
at the edge of time-reversal symmetric topological insulators
(the quantum spin Hall effect), 
Congjun Wu, B. Andrei Bernevig, and Shou-Cheng Zhang,
Phys.\ Rev.\ Lett.\ \textbf{96}, 106401(2006);
C. Xu and J. E. Moore, 
Phys. Rev. B {\bf 73}, 045322 (2006).





\bibitem{Shankar}
R.\ Shankar,
Phys.\ Rev.\ Lett.\
\textbf{46}, 379 (1981);
Phys.\ Rev.\ Lett.\
\textbf{50}, 787 (1983).


\bibitem{footnoteSO8}
While the interpretation
in terms SO(8) symmetry is elegant,
it is not entirely necessary to have
the same Fermi velocity for all eight (vector)
Majorana fermions.
The physics described below
has been discussed in the past
in the context of, among others,
the Hubbard model on the two-leg ladder,
which does not have, at least microscopically,
SO(8) symmetry,
{\it i.e.}, the Abelian bosonization can be used.




\bibitem{Scalapino1998}
D.\ Scalapino,
Shou-Cheng Zhang, 
and
W.\ Hanke, 
Phys.\ Rev.\ B
\textbf{58}, 443 (1998).


\bibitem{Lin1998}
H.\ -H Lin,
L.\ Balents,
and
M.\ P.\ A.\ Fisher, 
Phys.\ Rev.\ B
\textbf{58}, 1794 (1998).


\bibitem{Marston2002}
J.\ B.\ Marston,
J.\ O.\ Fjaerestad
and
A.\ Sudb\o,
Phys.\ Rev.\ Lett.\
\textbf{89}, 056404 (2002).


\bibitem{Fjaerestad2002}
J.\ O.\ Fjaerestad
and 
J.\ B.\ Marston,
Phys.\ Rev.\ B
\textbf{66}, 125106 (2002).


\bibitem{Tsuchiizu2002}
M.\ Tsuchiizu and A.\ Furusaki,
Phys.\ Rev.\ B
\textbf{66}, 245106 (2002).


\bibitem{Schollwock2002}
U.\ Schollw\"ock,
S.\ Chakravarty, 
J.\ O.\ Fjaerestad,
J.\ B.\ Marston,
and
M.\ Troyer,
Phys.\ Rev.\ Lett.\
\textbf{90}, 186401 (2003).


\bibitem{WuLiuFradkin2003}
C.\ Wu,
W.\ V.\ Liu,
and 
E.\ Fradkin, 
Phys.\ Rev.\ B
\textbf{68}, 115104 (2003).



\bibitem{Wu2003}
Congjun Wu,
Jiang-ping Hu,
and
Shou-cheng Zhang,
Phys.\ Rev.\ Lett.\
\textbf{91}, 186402 (2003).


\bibitem{Nonne2010}
H. Nonne, E. Boulat, S. Capponi, and P. Lecheminant,
Phys.\ Rev.\ B \textbf{82}, 155134 (2010).




\bibitem{latticevscont}
It is not a priori justified to take a lattice model
to discuss physics of an edge theory.
In fact,
from the modular invariance of the edge theory
when $N_f=8$,
we expect that the edge theory is ``trivial'' and
can be described by a lattice model
which can be gapped;
we are prefetching our discussion
in Secs.\
\ref{global gravitational anomaly}
and
\ref{edge theory of Z2xZ2 symmetric topological phase}.

\bibitem{qi2008}
Xiao-Liang Qi and Shou-Cheng Zhang,
Phys. Rev. Lett. \textbf{101}, 086802 (2008).

\bibitem{LevinStern2009}
Michael Levin and Ady Stern,
Phys. Rev. Lett. \textbf{103}, 196803 (2009).

\bibitem{RyuMooreLudwig2012}
Shinsei Ryu, Joel E. Moore, and Andreas W. W. Ludwig,
Phys.\ Rev.\ B \textbf{85}, 045104 (2012).

\bibitem{WnagQiZhang2011}
Zhong Wang, Xiao-Liang Qi, and Shou-Cheng Zhang,
Phys.\ Rev.\ B \textbf{84}, 014527 (2011).

\bibitem{Stone2012}
Michael Stone,
Phys. Rev. B \textbf{85}, 184503 (2012).

\bibitem{AlvarezGaumeWitten1983}
L.\ Alvarez-Gaum\'{e} and E.\ Witten,
Nucl.\ Phys.\ B \textbf{234}, 269 (1983).

\bibitem{Volovik90}
G. E. Volovik,
JETP Lett.
\textbf{51},
125,
(1990).

\bibitem{readgreen}
N. Read and Dmitry Green,
Phy.\ Rev.\ B
\textbf{61},
10267
(2000).

\bibitem{Cappelli01}
A.\ Cappelli,
M.\ Huerta,
and
G.\ R.\ Zemba,
Nucl.\ Phys.\ B \textbf{636}, 568 (2002).

\bibitem{Witten1985}
E.\ Witten,
Commum. Math. Phys. \textbf{100}, 197 (1985).


\bibitem{Polchinski98}
J.\ Polchinski,
\textit{String Theory},
Cambridge University Press (Cambridge, UK), (1998).

\bibitem{Ginsparg91}
Paul Ginsparg,
in
\textit{Fields, Strings and Critical Phenomena: Proceedings 
(Les Houches 1988)},
ed. by E.\ Brezin and Jean Zinn-Justin,
pp. 1-168.
Amsterdam: North-Holland (1990).

\bibitem{Imamura2011}
Y. Imamura,
{\it Basics of Superstring Theory},
Saiensu-Sha (Tokyo, Japan) (2011).

\bibitem{Cardy1986}
J. L. Cardy,
Nucl.\ Phys.\ B, \textbf{270}, 186 (1986).


\bibitem{Cappelli96}
A.\ Cappelli,
and 
G.\ R.\ Zemba,
Nucl.\ Phys.\ B \textbf{490} 595 (1997).



\bibitem{Cappelli09}
A.\ Cappelli,
G.\ Viola
and
G.\ R.\ Zemba,
Annals of Physics
\textbf{325}, 465 (2010).


\bibitem{Cappelli09b}
A.\ Cappelli,
and
G.\ Viola,
J.\ Phys.\ A \textbf{44}, 075401 (2011).


\bibitem{footnote1}
Furthermore,
something special happens when $N=4$ ($2N=8$):
the total partition function actually vanishes
$Z_L(\tau)=0$ identically,
because of 
the fact that $Z^1{ }_1(\tau)$ vanishes due to 
a fermion zero mode,
and of the ``abstruse identity'' by Jacobi.
This is a manifestation of spacetime supersymmetry.
The fact that the partition function accidentally vanishes
makes 
the meaning of achieving modular invariance 
somewhat obscure.  
However, the modular invariance/non-invariance
should be discussed also at the level of correlation functions
({\it i.e.}, with insertions of operators), and while the partition
function is zero, the correlation functions are not zero
in general.
In the situation at hand, the modular properties are the same
with or without insertion of operators.
\cite{Polchinski98}



\bibitem{footnoteIsing}
As before, there is a sign ambiguity $\pm$ here.  
However, 
we can actually change this sign by Kremers-Wannier duality --
so the sign convention is fixed once we fix
convention for the spin $\sigma$
and disorder $\mu$ operators.
\cite{Ginsparg91}


\bibitem{CappelliItzyksonZuber}
A.\ Cappelli,
C.\ Itzykson,
J.\ -B.\ Zuber,
Comm.\ Math.\ Phys.\
\textbf{113}, 1 (1987).

\bibitem{Kato1987}
A.\ Kato,
Mod.\ Phys.\ Lett.\ A
\textbf{2}, 585 (1987).

\bibitem{Nepert2011}
Titus Neupert, Luiz Santos, Shinsei Ryu, Claudio Chamon, and Christopher Mudry,
Phys.\ Rev.\ B \textbf{84}, 165107 (2011).


\bibitem{Kitaev2005}
A.\ Kitaev,
Ann.\ Phys.\ \textbf{321}, 2 (2006). 

\bibitem{QiHughesZhang2007}
Xiao-Liang Qi,
Taylor L. Hughes,
and
Shou-Cheng Zhang,
Nat. Phys. \textbf{4}, 273 (2008). 

\bibitem{Maciejko2009}
Joseph Maciejko, Chaoxing Liu, Yuval Oreg, Xiao-Liang Qi, Congjun Wu,
and Shou-Cheng Zhang,
Phys.\ Rev.\ Lett.\ \textbf{102}, 256803 (2009).













\end{thebibliography}
\end{document}